\documentclass[8pt]{article}

\usepackage[margin=1in]{geometry}  
\usepackage{graphicx}              
\usepackage{hyperref}
\usepackage{url}

\usepackage[font=small,format=plain,labelfont=bf,up,textfont=it,up]{caption}
\usepackage{subcaption}
\usepackage{amsmath}               
\usepackage{amsfonts}              
\usepackage{amsthm}                
\usepackage[font={small,it}]{caption}

\begin{document}
\nocite{*}
\title{Measuring Software Diversity, with Applications to Security}

\author{Julio Hernandez-Castro\\
jch27@kent.ac.uk\\
Department of Computer Science \\
The University of Kent \\
Canterbury CT2 7NF, Kent, UK\\\\
Jeremy Rossman\\
J.S.Rossman@kent.ac.uk\\
School of Biosciences \\
The University of Kent \\
Canterbury CT2 7NJ, Kent, UK}

\maketitle

\begin{abstract}
  In this work, we briefly introduce and discuss some of the diversity measures used in Ecology. 
  After a succinct description and analysis of the most relevant ones, we single out the Shannon-Weiner index. 
  We justify why it is the most informative and relevant one for measuring software diversity. 
  Then, we show how it can be used for effectively assessing the diversity of various real software ecosystems. 
  We discover in the process a frequently overlooked software monopoly, and its key security implications. 
  We finally extract some conclusions from the results obtained, focusing mostly on their security implications.
\end{abstract}

\section{Introduction}

\subsection{Some Biological Examples}

It has long been appreciated in Biology that monocultures are extremely vulnerable to pathogens.
One particularly notable example is the Irish potato Great Famine in the XIX century, caused by \emph{Phytophthora infestans}
and exacerbated by a disproportionate percentage of all potatoes grown in Ireland being from the same variety (Irish Lumper).

Another interesting example is the near extinction caused by the fungus \emph{Fusarium oxysporum} of the once very popular
Gros Michel banana. This variety was replaced commercially around 1950 by the nowadays omnipresent Dwarf Cavendish banana
variety, which was also turned into a massive monocultive. This plants are all clones (i.e genetically identical),
so they cannot evolve any resistance to new diseases. The parallelism with current software is striking. Unsurprisingly, the Cavendish clones, so popular in our supermarkets,
seem to be about to suffer a similar fate at the hands of a very related pathogen \emph{Panama disease, Fusarium oxysporum
Tropical Race 4} \cite{ploetz2006panama}.

\begin{figure}
\centering
\begin{subfigure}{.5\textwidth}
  \centering
  \includegraphics[width=.4\linewidth]{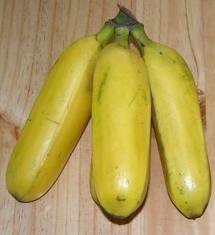}
  \caption{The Gros Michel banana.}
  \label{fig:sub1}
\end{subfigure}%
\begin{subfigure}{.5\textwidth}
  \centering
  \includegraphics[width=.4\linewidth]{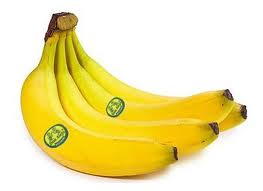}
  \caption{The Cavendish banana.}
  \label{fig:sub2}
\end{subfigure}
\caption{The Gros Michel banana is now almost extinct and no longer viable for commercial exploitation because of its massive popularity and consequent monoculture in the first half of the 20$^{th}$ century. The Cavendish banana, seriously threaten now by a variant of same pathogen that wroke havoc 50 years ago on the Gros Michel, is another example of the dangers of massive monoculture in the Biological World.}
\label{fig:test}
\end{figure}

\subsection{Diversity in Computer Science \& Security}
\label{sect:divincs}

This well studied phenomenon in Ecology, where many more examples than the two previously cited exist, has being observed also in Software
ecosystems. Attempts to extrapolate this concept to Software have been, however, unsucessful mostly due to the lack of a proper way to quantitatively measure diversity.

So far, the consequences and claims published in the literature were mostly based in
common sense and intuition. The lack of a quantitative way of measuring diversity hindered to some extend the credibility of the
arguments displayed, and made difficult if not impossible to compare different ecosystems to decide where more urgent action was
needed.

We try to address this problem in our work by briefly introducing some well-known diversity measures used in Biology and Ecology, and
then proposing what we believe is the most informative and useful one, called the Shannon-Wiener index. After doing so, we show how
it can be applied to a number of different Software ecosystems, and the security implications that can be extracted from the results
obtained.

Diversity in software was first proposed for safety, not security, in the context of fault tolerance \cite{randell1975system}. 
Then, N-version programming attempted to build functionally equivalent but completely different code by isolating $N$ independent teams of developers with a common 
specification \cite{chen1978n}. A majority vote is applied to the result, and in this way robustness is achieved against a low number ($<\frac{N}{2}$) of faults. 
The aim is, of course, to obtain failure diversity by this procedure, but unfortunately later experiments showed this was not the case \cite{knight1986experimental}, 
though improvements where obtained.

Forrest and her colleagues have also contributed significantly to apply concepts from Biology to Computer Security, explicitly stating \cite{forrest1997building} that diversity
could reduce the impact of any security weakness. They say: ``Diversity is an important source of robustness in biological systems'', and even more importantly for the aims of this paper ``A
stable ecosystem, contains many different species which occur in highly-conserved frequency distributions. If this diversity is lost and a few species become dominant, the
ecosystem becomes susceptible to perturbations such as catastrophic fires, infestations, and disease.'' They furthermore propose some general countermeasures to increase diversity, 
such as avoiding unnecessary consistency, adding or deleting non-functional code, code reordering and some memory allocation techniques. 
These and other similar ideas for automatically increasing diversity at a relatively low cost have been explored further: the two main research paths are based on memory rearrangement \cite{bhatkar2003address}\cite{xu2003transparent} and randomising the instruction set \cite{barrantes2003randomized}. Another promising approach, loosely based on the N-version concept, to inject automated diversity and provide detection and disruption against attacks, is N-variant systems \cite{cox2006n} \cite{nguyen2008security} \cite{weatherwax2009model}.

Software monocultures as a security threat have been explicitly studied before, most notably by Dan Geer in \cite{de2003monopoly} and Geer \emph{et al.} in \cite{geer2003cyberinsecurity}.
These two documents focused mainly on the risks to cybersecurity posed by the dominance of Microsoft's products. Geer, then Chief Technical Officer and founding member of AtStake, 
was fired by his employer for publishing the first report. AtStake was a supplier to Microsoft who, to put it mildly, did not like the research at all. Particularly assertions like this one ``Governments must set an example with their own internal policies and with the regulations they impose on industries critical to their societies. They must confront the security effects of monopoly and acknowledge that competition policy is entangled with security policy from this point forward'', or ``The threats to international security posed by Windows are significant, and must be addressed quickly''. Another notable early work on the dangers of monoculture is \cite{stamp2004risks}, where the author describes the impact of the W32/Blaster worm in August 2003, and concludes ``More diversity in the OS market would have limited the number of 
susceptible systems, thereby reducing the level of infection''.

Two relatively recent studies clearly show statistically significant evidence of the security benefits of using environment with very diverse software.
In a particularly insightful work \cite{garcia2013analysis} 18 years worth of vulnerabilities (2270) were investigated, targeting 11 different Operating Systems, 
and the results showed that large potential security gains could be attained from using diverse operating systems in a replicated intrusion-tolerant system. 
Authors presented several strategies for system designers to optimally choose the best combination of operating systems and, curiously enough, came with the same solution 
regardless of which type of vulnerability was more relevant for a particular system. This optimal combination was one including OpenBSD, Debian, Solaris and Windows 2003. 
That was proved to be the best possible combination of just four operating systems for an intrusion-tolerant configuration, for their particular dataset.

Another seminal piece of work on the effectiveness of software diversity for security is \cite{han2009effectiveness}.
It presents, similarly to the previous study, a systematic analysis of security vulnerabilities published in 2007. It focus particularly in using diversity for 
intrusion detection, and in measuring how effective is one of the cheapest diversity-increasing measure of them all: just combining different off-the-shelf software. 
The results are really encouraging and statistically significant, as they hold over the more than 6000 vulnerabilities published in 2007 they studied.


\subsection{Organisation of the Paper}

This paper is organised as follows: First, we briefly introduce in section \ref{sect:basics} a number of the most significant diversity measures used in Ecology.
Later, we criticise them following closely the arguments and reasoning of Lou Jost \cite{louweb:201:Misc}, and justify why we believe the modified
Shannon-Wiener is the best suited for our purpose. In section \ref{toyexamples} we show some examples of how this diversity measure can be applied, starting with some toy examples. Afterwards, in section \ref{sect:diversityofswecos} we use it in different real and current software ecosystems. We later analyse in detail the results obtained, particularly regarding their security implications.
We finish with some conclusions in section \ref{sect:conclusions} and present some biologically inspired ideas for further work in section \ref{sect:futureworks}.

\section{Diverse Diversity Measures}
\label{sect:basics}

\subsection{Ecological Diversity Measures}
\label{toyexamples}

According to Lou Jost \cite{louweb:201:Misc}, a well-known ecologist specialised in diversity, the main diversity indexes that have
been used in Biology in the past have been quite misleading. They led on a number of occasions to wrong conclusions and
erroneous interpretations of the data. Often-times researchers came to the depressing conclusion that diversity could not be
measured. As Jost puts it: ``The plethora of diversity indices and their conflicting behavior has led some authors [...] to conclude
that the concept of diversity is meaningless''. Another undesirable consequence is that the results of different researchers,
favouring different indexes, where frequently impossible to compare.

Jost describes the whole problem quite lucidly: ``The radius of a sphere is an index of its volume but is not itself the volume, and
using the radius in place of the volume in engineering equations will give dangerously misleading results. This is what biologists
have done with diversity indices.''

He considers in his seminal work \cite{jouoikos} entropic measures such as the Gini-Simpson, the Shannon-Wiener, the Renyi,
the Tallis/HCDT and other related concepts such as species richness, Simpson concentration and similar ones. He criticises many
of these indexes as not being diversities at all, and
claims these confusions have lots of practical implications.

For trying to address this problem, he focuses on a very simple population composed of $S$ equally common species, and observes that for
most of these indexes we will not obtain the most natural diversity value of $S$ but something ranging from $log(S)$ to $1/S$ or
even $1-1/S$. Some will be confined to a value in the $[0,1]$ interval, while others can range up to $\infty$. He strongly argues
that all indexes cannot be called diversities and treated interchangeably. He additionally poises that their mathematical
characteristics do not reflect at all the intuition that Biologists have of the concept of 
diversity\footnote{The properties intuitively expected of a diversity by Jost are two: 1. It always returns $S$ when applied to a community with $S$ equally-common species, and 2. It 
possess the ‘‘doubling’’ property, i.e. : suppose we have a community
of $S$ species with arbitrary frequencies $p_1, ... p_i, ... p_s$, with diversity $D$. Suppose we divide each species into two equal groups, say males and
females, and we treat each group as a separate ‘‘species’’. Intuitively, we have doubled the diversity of the community by this reclassification, and indeed the diversity of
the doubled community should be $2D$.}. For solving these problems
and to put different diversity indexes in an equal footing, he proposed to turn them into what he calls true diversities. 
He offers a list of mathematical transformations that can convert each of them into a proper diversity measure, with all the required properties.

The transformations will change any of the measures into real diversities, and he suggests that the common unit of
species diversity should be called the \emph{effective number of species}. 

After applying the proposed transformations, these different
values will behave as intuitively expected\footnote{For example, in a community where all species are equally common, diversity
should be proportional to the number of species. This very intuitive property is not met by the Shannon-Weiner entropy before the
proposed conversion into a true diversity, but it holds after it.}, and become fully meaningful.

Of all the studied indices, the Shannon-Wiener seems to be, for a number of mathematical reasons detailed in \cite{jouoikos}, the
most adequate. It is even called by Jost the ``most profound and useful of all diversity indices''. It has also the advantage to be
intimately related to the well-known concept of entropy in Information Theory. The problem is that, as it stands in equation
\ref{eq:shannonentropy}:

\begin{equation} \label{eq:shannonentropy}
H(S) = -\sum_{i=1}^{S}{p_i \ln(p_i)}.
\end{equation}

it is not a true diversity.\\

Fortunately, it is quite easy to convert it into one, by applying exponentiation as in the following
equation:

\begin{equation} \label{eq:shannondiversity}
D(S)=\exp^{H(S)} = \exp^{-\sum_{i=1}^{S}{p_i \ln(p_i)}}.
\end{equation}

Of course, equation \ref{eq:shannondiversity} can be trivially simplified into

\begin{equation} \label{eq:shannondiversitysimplified}
D(S)=\exp^{H(S)} = \exp^{-\sum_{i=1}^{S}{p_i \ln(p_i)}}=\prod_{i=1}^{S}{\frac{1}{p_i^{p_i}}}.
\end{equation}

but curiously enough, we did not find this straightforward transformation in the original paper.\\

So the units of this diversity value $D$ are \emph{equivalent species number}\footnote{We will abbreviate them as \emph{\textbf{esn}} in the following.}
and they admit an easy and intuitive interpretation as the number of equally distributed species that will achieve an equivalent diversity.

Let us see an extremely simple example:

\begin{center}
\begin{tabular}{|c||c|c|c|c|}
\hline
$Species$ & 1 & 2 & 3 & 4 \\
\hline
Community A & 0.25 & 0.25 & 0.25 & 0.25 \\
\hline
Community B & 0.25 & 0.75 & 0 & 0 \\
\hline
Community C & 0.1 & 0.2 & 0.3 & 0.4 \\
\hline
\end{tabular}
\end{center}

Following equation \ref{eq:shannondiversity}, we have that

$$H(A) = -\sum_{i=1}^{4}{\frac{1}{4} \ln(\frac{1}{4})}=-\ln(\frac{1}{4})=\ln(4)$$
$$D(A) = \exp^{\ln(4)}= 4$$\\ or, alternatively
$$D(A) = \prod_{i=1}^{S}{\frac{1}{p_i^{p_i}}} = ((4)^{\frac{1}{4}})^4=4$$\\

So we have the quite intuitive value of 4 \emph{\textbf{esn}} (effective number of species), which is the maximum attainable with 4 species.\\

Similarly,

$$H(B) = -\sum_{i=1}^{2}{p_i \ln(p_i)}= -\frac{1}{4} \ln(\frac{1}{4}) - \frac{3}{4} \ln(\frac{3}{4}) = 0.34657 + 0.21576 = 0.56233$$
$$D(B) = \exp^{0.56233}= 1.75476$$\\ or, alternatively
$$D(B) = \prod_{i=1}^{S}{\frac{1}{p_i^{p_i}}} = 4^{\frac{1}{4}} \cdot (\frac{4}{3})^{\frac{3}{4}}=1.75476$$\\

It is interesting to note that Community B is significantly less diverse that Community A, where diversity is maximal. Also, it
is quite intuitive that, as there are only 2 species in Community B its diversity should be upper bounded by 2 \emph{\textbf{esn}}. A further
conclusion would be that, as these two species are not equidistributed, the diversity should be lower than 2 \emph{\textbf{esn}}. The computed result
of roughly 1.75 \emph{\textbf{esn}} is consistent with all these intuitions.

As a last example, let us check what happens in Community C. Without a diversity measure as the one at hand, it would be not
easy to compare the diversities of Communities B and C. Thanks to this concept, this can be done effortlessly:


$$D(C) = \prod_{i=1}^{S}{\frac{1}{p_i^{p_i}}} = 10^{\frac{1}{10}} \cdot 5^{\frac{1}{5}} \cdot {(\frac{10}{3})}^{\frac{3}{10}} \cdot (\frac{5}{2})^{\frac{2}{5}}=(\frac{5}{3^{\frac{3}{10}}})=3.59611$$\\

The result obtained seems to confirm the intuitive idea that the diversity of Community C is much higher that that of
Community B. In fact, it is not that far away from the maximal one of 4 \emph{\textbf{esn}} observed in Community A.

\section{Diversity of Some Software Ecosystems}
\label{sect:diversityofswecos}

For seeing how the diversity measure that we just introduced can be applied to real life scenarios, not only to the toy examples shown in Section \ref{toyexamples},
we will explain in the following our analysis of two particular ecosystems that surprised us for their extremely poor diversity. 
We have additionally included another interesting ecosystem for comparison's sake.

\subsection{Desktop Operating Systems}
\label{sect:pcs}

It will hardly come as a surprise to anyone, particularly after the efforts of Geer \emph{et al.} in \cite{de2003monopoly}, but we have been able to confirm their finding and, above all,
to put a figure to the intuitive quite clear fact that this ecosystem suffers from very poor diversity.

\begin{figure}[ht]
\begin{center}
\includegraphics[scale=0.5]{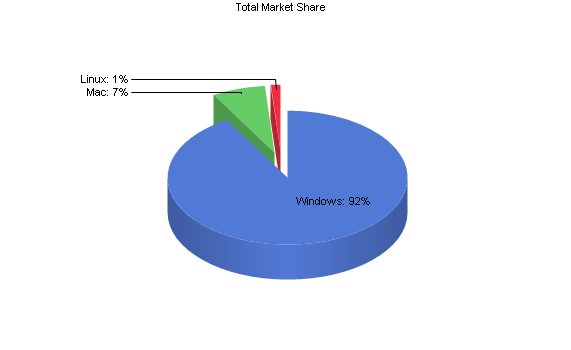}
\caption{The Desktop Operating System market, as measured by NetMarketShare in June 2013.}
\end{center}
\label{fig:pcs}
\end{figure}

Let us analyze these results in greater detail. They are shown in Table \ref{tablepc} below:

\begin{center}
\begin{tabular}{|c||c|}
\hline
Operating System & Market Share \\
\hline
Windows & 91.51\% \\
\hline
Mac & 7.20\% \\
\hline
Linux & 1.28\% \\
\hline
\end{tabular}
\label{tablepc}
\end{center}

Applying our proposed diversity measure we get that the diversity for this ecosystem is just 1.386 \emph{\textbf{esn}}.
This is an extremely small figure that is never found in any living ecosystem, not even in those heavily altered by farming.
This is also a clear indication of the enormous rewards that can be extracted from investing in more complex and sophisticated attacks in this ecosystem.
All in all, very worrying though not unexpected. This fully justifies the work of Geer \emph{et al.} but also shows very little has changed in the last 10 years.

Many would dispute the figure of 1.386 \emph{\textbf{esn}}, arguing that this classification of all Desktop operating systems into three groups is too coarse and misses important
details. This may be partly true, so we wanted to investigate the other end of the spectrum, when we (wrongly!) assume that different versions of different
desktop operating systems are completely unrelated. This scenario is show below:

\begin{center}
\begin{tabular}{|c||c|}
\hline
Operating System & Market Share \\
\hline
Windows 7 & 44.37\% \\
\hline
Windows XP & 37.17\% \\
\hline
Windows 8 & 5.10\% \\
\hline
Windows Vista & 4.62\% \\
\hline
Mac OS X 10.8 & 3.14\% \\
\hline
Mac OS X 10.6 & 1.76\% \\
\hline
Mac OS X 10.7 & 1.73\% \\
\hline
Linux & 1.28\% \\
\hline
Mac OS X 10.5 & 0.43\% \\
\hline
Windows NT & 0.19\% \\
\hline
Mac OS X 10.4 & 0.10\% \\
\hline
Windows 2000 & 0.04\% \\
\hline
Mac OS X 10.9 & 0.02\% \\
\hline
Mac OS X NVR & 0.02\% \\
\hline
Windows 64 & 0.01\% \\
\hline
\end{tabular}
\label{tablepcindetail}
\end{center}

\begin{figure}[ht]
\begin{center}
\includegraphics[scale=0.5]{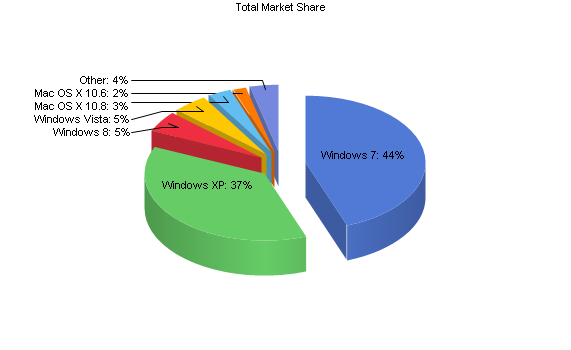}
\caption{The Desktop Operating System market by OS version, according to NetMarketShare in June 2013}
\end{center}
\label{fig:pcsindetail}
\end{figure}

With the figures shown in Table \ref{tablepcindetail}, the diversity of this ecosystem is 3.971 \emph{\textbf{esn}}.
This is still too low in comparison with living ecosystems, and a clear demonstration of the poor diversity of this market.
Of course, the real figures will lay somewhere in between these two values of 1.386 and 3.971 \emph{\textbf{esn}}, possibly much 
closer to the lower due to millions of common lines of code between many of the OS that we treat here as different.
In any case, these figures reveal that added diversity is much needed in this ecosystem for increased security.

\subsection{Supercomputers}
\label{sect:supercomputers}

A much more surprising finding in the course of our research is that the Desktop Operating System market is not the only with a worryingly
low diversity, and that it is not even the worst offender.

\begin{figure}[ht]
\begin{center}
\includegraphics[scale=0.6]{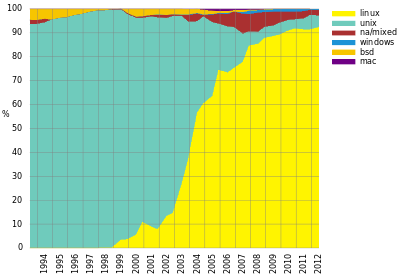}
\caption{OS evolution of Top 500 supercomputers in the World on June 2013, from http://www.top500.org} \label{supercomputers}
\end{center}
\end{figure}

The values shown in figure \ref{supercomputers} lead to a shocking diversity measure of 1.269 \emph{\textbf{esn}}, even worse that the one for the Desktop PC operating systems.

Recent news \cite{Goodin2013} have shown the first publicly disclosed attack against supercomputers. Most of these have supposedly 
very tough access control policies and high security levels, so it is possible that miscreants have also discovered the vulnerability of this ecosystem and are increasingly 
targeting it, particularly now that access to massive computational power can be cashed-in almost instantly and with some anonymity thanks to cryptocoins such as Bitcoin 
and LiteCoin.
We predict much more attacks of this type in the future, unless something is done to revert the extreme monoculture highlighted by our work, 
that translates into very poor diversity and hence extreme vulnerability due to the enormous attraction of massive impact and replication of a single weakness or exploit.

Most authors in the past have covered the security implication of Microsoft's monopoly \cite{de2003monopoly}\cite{geer2003cyberinsecurity}\cite{stamp2004risks}, but very few have
extended these critics to other markets. We are the first to highlight another crucial environment (very attractive for intellectual property theft, scientific knowledge stealing, 
cryptocoin mining, etc.) which has even lower diversity, and to ask for direct and quick measures to be put in place to protect this critical infrastructure.

\subsection{Mobile and Tablet Operating System}
\label{sect:mobile}

Another well-known market where poor diversity has been repeatedly been cited as one of the main causes of its multiple security problems is that of mobile and tablet operating systems.

\begin{figure}[ht]
\begin{center}
\includegraphics[scale=1.0]{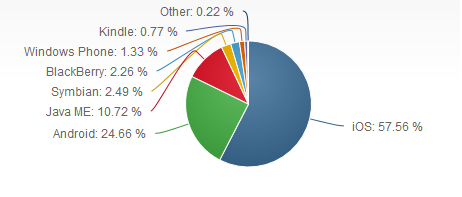}
\caption{The Mobile and Tablet Operating System market in June 2013, by NetMarketShare.}
\end{center}
\label{mobile}
\end{figure}

The figures obtained from NetMarketShare on this ecosystem, schematically shown in Figure 5, lead to a diversity of 3.29 \emph{\textbf{esn}} at the OS level 
and 7.305  \emph{\textbf{esn}} at the version level. 
As in the case analysed in \ref{sect:pcs}, the real diversity will lie somewhere between these two values, probably much closer to the lower end due to the enormous similarities (and hence vulnerability exposure) of Android versions 1.6, 2.1, 2.2, 2.3, 3.1, 3.2, 4.0, 4.1, 4.2 and the iOS versions used in the iPad, iPhone and iPod, that are all, quite unrealistically, treated as unrelated in the version level analysis.

This diversity is, anyway, really poor by any of the standars found in natural ecosystems, but relatively high if compared with the rest of pathologically poor markets analysed 
in this paper.

\section{Conclusions}
\label{sect:conclusions}

 There is a remarkable parallelism between massive farming and the fact that hundreds of millions of exact software clones (OS, databases, webservers, etc.) exist across the Internet. Due to this monoculture, they offer a particularly attractive target that can potentially lead to tens of millions of compromised machines. And that with only a single vulnerability or exploit. The analogy with the Irish potato or the Gros Michel/Cavendish banana is clear. This lack of diversity can be used to create large botnets that generate huge profits for criminals by launching DDoS attacks, mining cryptocoins, etc.
 But many other software ecosystems suffer from a lack of diversity that, thanks to our work, can only now be measured and compared quantitatively. 
 
 We have confirmed that a very well known and large market (that of Desktop OS) has indeed very low diversity, fully supporting the arguments of Geer et al. 
 But  we have also discovered a frequently forgot ecosystem (that of the top 500 supercomputers\footnote{These are very attractive targets. Access to them will grant privileged knowledge of confidential scientific and even military data. It can even allow attackers to tamper with the result of scientific and military experiments, inducing rivals to abandon or revise theories and gaining valuable intelligence and commercial insights.}) that has even worst 
 characteristics, and could lead to very serious security compromises of some of the most powerful computers on Earth. 
 There is some evidence that attacks aiming at these supercomputers have recently taken place, possibly taking advantage (the criminal boasted having obtained access
 to almost half of the top 500 supercomputers) of their extreme lack of diversity. Claims like that become, unfortunately, quite believable thanks to its very poor diversity.
 This top-500 ecosystem is curious and quite interesting in itself. Many authors have previously blamed Microsoft's monopoly aspirations and its blind desire for profit as 
 one of the leading explanations for the lack of diversity in desktop OS environments. This narrative, of course, does not apply to the top-500 market where the dominating OS is 
 a free open source one. We have to seek other reasons for this undesirable state of affairs, and most likely apply different measures to correct it.

 We believe urgent actions should be taken to force a significant increase in the diversity of critical software ecosystems. One could argue that this is by no means straightforward to achieve (particularly in the OS Desktop market) as economic offer and demand forces act over it. This is only partly true and, as requested by Geer \emph{et al.} in \cite{geer2003cyberinsecurity}, governments could do much more to adapt their software and hardware contracts (some of these including blocks of millions of machines) to explicitly increase their overall diversity to acceptable levels. The role of a correct government policy could be even greater in niche ecosystems such as the top-500, where many of the supercomputers are in full or in part government-owned.

 Increased diversity will likely stop the now common phenomenon on botnets with millions of compromised machines, significantly reducing their size and increasing the investment wrongdoers have to put to herd so many machines, seriously limiting their profits.

 Of course, higher diversity should ideally be implemented by automatic means such as those mentioned in Section \ref{sect:divincs}. Advances in this area will have a high impact for improving the overall security of the Internet. But it is also true that managing different version of the same software could imply some extra costs in terms of maintenance and support. Fortunately, as shown in \cite{han2009effectiveness} and \cite{garcia2013analysis}, there are much cheaper options that actually work quite well by simply increasing he diversity of the main off-the-shelf components of a software system. Moves in this direction will very beneficial for security and will guarantee less exposure to vulnerabilities, while other techniques such as address space layout randomization, randomisation of the instruction set, N-variants and others \cite{bhatkar2003address}\cite{xu2003transparent}\cite{barrantes2003randomized}\cite{cox2006n}\cite{nguyen2008security}\cite{weatherwax2009model} are fully deployed.

\section{Future Works}
\label{sect:futureworks}

We acknowledge the limitations of the techniques we show in this work to predict exposure to cybersecurity attacks, particularly because diversity is a measure taken at a certain time and, as an isolated photography, does not tell the whole story. That said, our findings also point out towards the fact that no real life ecosystems (not even massive farms) would have such a low diversity as the ones encountered in this work.

One possible next step could be to use this new capability of precisely measuring diversity to, in a similar way to how it is done for other species, compute the probability of survival of a given software ecosystem. Biologists use for that some simple models and run multiple simulations that try to predict the evolution of each species, and see how many of them survive for how long in how many simulations. There is no reason to believe this could not be also done for software ecosystems, possibly with different models, and we believe exploring this path of linking diversity and extinction probability is a worthy pursuit.

We will also use the new presented quantitative measure of diversity for studying how different software ecosystems evolve on a yearly basis and, possibly, 
trigger alarms when diversity levels drop below certain thresholds. Establishing these thresholds with certainty, with the help of more powerful models and more input from biologists
and ecologists would be another interesting research objective.

Another interesting research direction will be to study how to reflect in our computations the similarity between different software components. 
One possibility would be to further establish analogies with biological ecosystems and their kinships as measured by genetic similitude. 
Common lines of code could be used as an initial consanguinity factor to more accurately measure diversity. In many cases this is a proprietary secret 
(How many lines of code have Windows 7 and XP?) but in others (Android, Linux) this data is readily available or relatively easy to obtain. Adapting the mathematical model proposed in this work to take into account this code and vulnerability similarities would also be a very interesting research avenue.

\section*{Acknowledgements}

The authors want to thank Dr. David Roberts, from the Durrell Institute of Conservation \& Ecology, School of
Anthropology \& Conservation, University of Kent, for invaluable discussions and help on the topic of biodiversity.

\bibliographystyle{plain}
'
\bibliography{paper}

\end{document}